\documentclass[prb,twocolumn,amsmath,amssymb,floatfix]{revtex4}
\usepackage{graphicx}

\newcommand{\be}{\begin{eqnarray}}
\newcommand{\ee}{\end{eqnarray}}
\newcommand{\bfr}{{\bf r}}

\newcommand{\tlA}{\tilde{A}}
\newcommand{\tlB}{\tilde{B}}
\newcommand{\tlP}{\tilde{P}}

\newcommand{\wbe}{\begin{widetext}}
\newcommand{\wee}{\end{widetext}}
\newcommand{\oncite}{\onlinecite}

\begin{document}
\draft

\title{Pseudo-potential of a power-law decaying interaction
in two-dimensional systems}

\author{Sheng-Min Shih$^{(1),(2)}$ and Daw-Wei Wang$^{(2),(3)}$}

\address{$^{(1)}$ Department of Physics, University of California, Berkeley, CA 94720
\\
$^{(2)}$ Physics Department, National Tsing-Hua University, Hsinchu 300, 
Taiwan
\\
$^{(3)}$ Physics Division, National Center for Theoretical Sciences,
Hsinchu 300, Taiwan}

\date{\today}

\begin{abstract}
We analytically derive the general pseudo-potential operator
of an arbitrary isotropic interaction for particles 
confined in two-dimensional (2D) systems, using the frame work
developed by Huang and Yang for 3D scattering. We also analytically derive 
the low energy dependence of the scattering phase-shift 
for an arbitrary interaction with a power-law decaying tail, 
$V_{\rm 2D}(\rho)\propto \rho^{-\alpha}$ (for $\alpha>2$).
We apply our results to the 2D dipolar gases ($\alpha=3$)
as an example, calculating the momentum and dipole moment
dependence of the pseudo-potential for both $s$- and $p$-wave 
scattering channels if the two scattering particles are in the same 2D
layer. Results for the $s$-wave scattering between
particles in two different (parallel) layers are also investigated. Our results can 
be directly applied to the systems of dipolar atoms and/or polar molecules in 
a general 2D geometry.
\end{abstract}


\maketitle
\underline{\it Introduction:}
Low-dimensional strongly correlated systems have been one
of the most important subjects in condensed matter physics in the
last few decades. From the many-body point of view, the
standard meanfield approximation for the 3D system
is usually broken down by thermal fluctuation at finite temperature, while
from the two-body point of view, the widely used first Born approximation for 
a 3D weak potential totally fails in lower dimensional systems
[\oncite{Landau}]. As a result, a proper effective
interaction (or called pseudo-potential) becomes essential to go beyond
the weak interaction limit of 3D scattering or in the long-energy limit of 
2D and 1D system. 
For the 2D system that we want to concentrate in this paper, an important progress
was made by Schick [\oncite{schick}] for the studying of 2D bosons with
a hard-disk potential, and the relevant applications to cold atom systems are also 
investigated by several groups [\oncite{Petrov,2D_pseudo,review}]. Recently
Kanjilal and Blume [\oncite{Blume}] further derived a general form of the 
pseudo-potentials for all angular momentum channels of a short-ranged interaction.
The pseudo-potential for a low-ranged dipolar interaction in 2D
systems was also studied in the $s$-wave channel by one of us [\oncite{bilayer_wang}],
but its extension to higher angular momentum channels is still unexplored, even 
though several important results have been carried out in 3D systems recently 
[\oncite{3D_dipole}].

In this paper, we systematically generalize earlier results and apply 
to the systems of 2D dipolar gases: (1) First, we apply Huang and
Yang's theory [\oncite{yang,note}] to derive a general form of the 
pseudo-potential in 2D systems. Our results can be shown equivalent
to Kanjilal and Blume's results [\oncite{Blume,private}], derived from 
another approach. (2) For a general power-law decaying 
interaction, $V_{\rm 2D}(\rho)\propto\rho^{-\alpha}$ ($\alpha>2$),
we further analytically calculate 
the low energy dependence of the scattering phase shift, upto a single non-universal
parameter to be determined by the short-ranged details of $V_{\rm 2D}$.
(3) Finally we apply our results to the study of $s$- and $p$-wave scattering
channels of dipolar interaction ($\alpha=3$), and numerically evaluate the non-universal
parameter for a model interaction. The $s$-wave scattering for the 
two scattering particles confined in two different (parallel) 2D layers
are also investigated, showing a Feshbach-like resonance even at zero dipole
moment limit. Our results can therefore be applied to the many-body physics 
of magnetic dipolar atoms [\oncite{Cr}], cold polar molecules
[\oncite{molecules}], or indirect excitons in a semi-conductor based
double-well system [\oncite{exciton}].

\underline{\it General pseudo-potential for 2D scattering:}
We start from solving the two-particle scattering problem of the 
following 2D Schr\"{o}dinger equation with total energy $E$:
\be
-\frac{\hbar^2}{2\mu}\nabla^2_\perp\psi(\bfr_\perp)
+V_{\rm 2D}(\rho)\psi(\bfr_\perp)
&=&E\psi(\bfr_\perp),
\label{eq_2body}
\ee
where $\nabla^2_\perp\equiv \frac{1}{\rho}\frac{\partial}{\partial\rho}
\rho\frac{\partial}{\partial\rho}+\frac{1}{\rho^2}
\frac{\partial^2}{\partial\phi^2}$ in cylindrical coordinate,
$\mu$ is the reduced mass, 
and $\psi(\bfr_\perp)$ is the scattered wavefunction in the relative
coordinate, $\bfr_\perp\equiv (x,y)$.
$V_{\rm 2D}(\rho)$ is the effective 2D interaction,
obtained by integrating out the transverse degree of freedom ($z$),
and is assumed to be isotropic about the $z$-axis here.
Note that we also have assumed that the transverse 
confinement potential is so strong that no confinement-induced 
resonance [\oncite{Petrov}] has to be considered here.

Since $V_{\rm 2D}(\rho)$ is assumed to be isotropic and decays
faster than $\rho^{-2}$ in large $\rho$, the wavefunction,
$\psi(\bfr)$, can be always expanded by noninteracting
eigenstates in large distance:
$\psi(\rho,\phi)=\sum_{m=0}^\infty u_m(k,\rho)
\sum_{\sigma=\pm}C_m^\sigma(k)e^{i\sigma m\phi}$, 
where $k=\sqrt{2\mu E/\hbar^2}$, and
$u_m(k,\rho)\equiv A_m(k)J_m(k\rho)+B_m(k)N_m(k\rho)$ is the radial
wavefunction with $J_m(x)/N_m(x)$ being 
the Bessel function of the first/second kind.
Here $A_m(k)$, $B_m(k)$, and $C_m^\pm(k)$ are coefficients to be determined 
by boundary conditions. Similar to the 3D case [\oncite{yang}],
we first investigate the short-distance behavior of such noninteracting
solution in the leading order terms:
\be
u_0(k,\rho)&\sim&A_0(k)+\frac{2B_0(k)}{\pi}
\ln\left(\frac{k\rho}{2\beta_0}\right),
\label{Phi_long_r0}
\\
u_m(k,\rho)&\sim&\frac{A_m(k)}{m!}
\left(\frac{k\rho}{2}\right)^m-\frac{B_m(k)(m-1)!}{\pi}
\left(\frac{2}{k\rho}\right)^m
\nonumber\\
&&+\frac{2B_m(k)}{\pi m!}\left(\frac{k\rho}{2}\right)^m
\ln\left(\frac{k\rho}{2\beta_m}\right).\ \ (m\neq 0)
\label{Phi_long_rm}
\ee
Here we define $\beta_m\equiv e^{-\gamma+H_m/2}$ with 
$\gamma\approx 0.57722$ being the Euler's constant and
$H_m\equiv\sum_{k=1}^m k^{-1}$ being the Harmonic 
number [\oncite{math}] (here $H_0\equiv 0$).
We note that the third term in the right hand side of 
Eq. (\ref{Phi_long_rm}), resulted from
the irregular solution, $N_m(k\rho)$, is 
of the same order (upto a logarithmic function) as the first term
if $A_m(k)\sim B_m(k)$. For a typical short-range interaction, 
however, this term can be neglected because 
$B(k)/A(k)\propto k^{2m}$ in the long wavelength limit. 
Here, since we want to derive a general
pseudo-potential for a power-law
decaying potential (see below) at a small (but finite) scattering energy,
we will still keep this term for the most general application.
Such hybridization between the regular and irregular solutions
of noninteracting partial waves does not 
exist in the 3D case [\oncite{yang}].
All other terms can be shown irrelevant to the derivation
of the pseudo-potential below.

To derive the proper pseudo-potential, 
we have to apply the noninteracting Hamiltonian on 
the asymptotic wavefunction above [\oncite{yang,note}], and integrating
over a small spherical area of radius $\rho$ by using Green' theorem [\oncite{note}]. 
Separating contributions from the $s$-wave and the 
non-$s$-wave parts, we obtain
\be
&&-\frac{\hbar^2}{2\mu}\left[\nabla^2+k^2\right]\psi(\bfr_\perp)
=-\frac{\hbar^2}{2\mu}\delta(\bfr_\perp)\left[4 B_0(k)\right.
\nonumber\\
&&\hspace{0.5cm}\left.+\sum_{m=1}^\infty
\frac{B_m(k)2^{m+2}m!}{(k\rho)^m}
\sum_{\sigma=\pm}C_m^\sigma(k)e^{i\sigma m\phi}\right].
\label{eq_B}
\ee
The next step is to rewrite the right hand side to be a function
of ${\cal P}_m(k)\equiv B_m(k)/A_m(k)$, which is the 
only quantity related to the phase shift
of the $m$th partial wave, $\delta_m(k)$. (In fact, ${\cal P}_m(k)
=-\tan\delta_m(k)$). In order to get an expression of 
$A_m(k)$, we have to take certain derivatives 
on the wavefunction and let $\rho\to 0$, as in the 3D case [\oncite{yang}]. 
After some simple calculation we obtain
\be
A_0(k)&=&\lim_{\rho\to 0}-\left(\ln\frac{k\rho}{2\beta_0}\right)^2\rho
\frac{\partial}{\partial\rho}\left[\frac{u_0(k,\rho)}
{\ln(k\rho/2\beta_0)}\right],
\label{eq_A0}
\\
A_m(k)&=&-\frac{2B_m(k)}{\pi}\left[H_{2m}
+\ln\left(\frac{k\rho}{2\beta_m}\right)\right]
\nonumber\\
&&+\lim_{\rho\to 0}\frac{m!}{(2m)!}\left(\frac{2}{k}\right)^m
\frac{\partial}{\partial\rho^{2m}}\left[\rho^m u_m(k,\rho)\right],
\label{eq_Am}
\ee
where we have used the following identity:
$\frac{\partial^{2m}}{\partial x^{2m}}
\left(x^{2m}\ln(x/b)\right)=(2m)!(H_{2m}+\ln(x/b))$ [\oncite{math}].
As a result, by combining Eqs. (\ref{eq_B}), (\ref{eq_A0}) and 
(\ref{eq_Am}), we find 
\be
-\frac{\hbar^2}{2\mu}\nabla^2\psi(\bfr_\perp)+
\sum_{m=0}^\infty \hat{\cal V}_{m}\psi(\bfr_\perp)=E\psi(\bfr_\perp),
\label{pseudo-eq}
\ee
where the pseudo-potential operator, $\hat{\cal V}_m$, is 
\be
\hat{\cal V}_{0}&\equiv &\delta(\bfr_\perp)
\frac{-4\hbar^2}{2\mu}{\cal P}_0(k)
\left(\ln\frac{k\rho}{2\beta_0}\right)^2\rho\frac{\partial}{\partial\rho}
\frac{1}{\ln(k\rho/2\beta_0)}
\label{V_ps0}
\\
\hat{\cal V}_{m}&\equiv&\delta(\bfr_\perp)
\frac{\hbar^2}{2\mu}\frac{4(m!)^2/(2m)!}
{{\cal P}_m(k)^{-1}+\frac{2}{\pi}\left[H_{2m}+\ln(k\rho/2\beta_m)\right]}
\nonumber\\
&&\times\frac{2^{2m}}{k^{2m}\rho^m}
\frac{\partial}{\partial\rho^{2m}}\rho^m.
\label{V_psm}
\ee
Eqs. (\ref{pseudo-eq})-(\ref{V_psm}) can be interpreted 
as the effect equation of Eq. (\ref{eq_2body}) with the same 
boundary condition at origin ($\bfr_\perp=0$) in the low energy limit 
($E,k\rho\to 0$), and hence is the 2D version of
Huang and Yang's result in Ref. [\oncite{yang}] (see Eq. (12)
therein). We note that one can show that the pseudo-potentials derived
above are equivalent to results of earlier work both in the $s$-wave channel
[\oncite{schick,Petrov,2D_pseudo,review,Blume}] and in the higher
angular momentum channels [\oncite{Blume,private}]. 

\underline{\it Pseudo-potential for $\rho^{-\alpha}$ interaction:}
After deriving the most general form of pseudo-potential for
2D scattering, we further derive an analytical closed form of the 
momentum dependence of ${\cal P}_m(k)$ for a power-law decaying potential:
$V_{\rm 2D}(\rho)\approx U/\rho^\alpha$ as $\rho\to\infty$. 
Here $U$ measures the strength of interaction, and $\alpha>2$ is the 
decay exponent. We start from the zero energy scattering ($E=k=0$) of 
Eq. (\ref{eq_2body}) and the radial wavefunction, $u_m(0,\rho)=u_m(\rho)$, 
can be calculated analytically: $u_m(\rho)=\tlA_m I_{\frac{m}{\xi}}
\left(\frac{\Delta_\alpha^\xi}{\xi\rho^\xi}\right)
+\tlB_mK_{\frac{m}{\xi}}\left(\frac{\Delta_\alpha^\xi}
{\xi\rho^\xi}\right)$ with $\tlA_m$ and $\tlB_m$ being the
coefficients to be determined by the short-distance behavior 
of $V_{\rm 2D}(\rho)$. Here $\xi\equiv \alpha/2-1$, $\Delta_\alpha\equiv
\left(\frac{MU}{\hbar^2}\right)^{\frac{1}{2\xi}}$,
and $I_m(x)/K_m(x)$ is the modified Bessel function of the first/second kind.
In the limit of long distance (or weak interaction, 
$\Delta_\alpha^\xi/\xi\rho^\xi\ll 1$), we have
\be
u_0(\rho)&\sim&\tlA_0+\tlB_0\left[\ln(2\beta_0\xi)+
\xi\ln\left(\frac{\rho}{\Delta_\alpha}\right)\right]
\nonumber\\
u_m(\rho)&\sim & \frac{\tlA_m(2\xi)^{-\frac{m}{\xi}}}
{\Gamma\left(\frac{m}{\xi}+1\right)}
\frac{\Delta_\alpha^m}{\rho^m}
+\tlB_m\frac{(2\xi)^{\frac{m}{\xi}}}{2}\Gamma\left(\frac{m}{\xi}\right)
\frac{\rho^m}{\Delta_\alpha^m},
\nonumber
\ee
which should be also reproducible 
by taking the zero energy limit ($k\to 0$) of 
Eqs. (\ref{Phi_long_r0}) and (\ref{Phi_long_rm}) (the last 
term of Eq. (\ref{Phi_long_rm}) can be neglected in this limit).
Therefore the relationship between ${\cal P}_m(k)$ and 
$\tilde{P}_m\equiv\tlB_m/\tlA_m$ can be easily derived to be
\be
{\cal P}_0(k)&=&\frac{\xi\pi/2}{\tlP_0^{-1}
+\ln(2\beta_0\xi)-\xi\ln(k\Delta_\alpha/2\beta_0)}
\label{P0_alpha}
\\
{\cal P}_m(k)&=&\frac{-2\pi(2\xi)^{-\frac{2m}{\xi}}}{m!(m-1)!}
\frac{(k\Delta_\alpha/2)^{2m}}{\Gamma\left(\frac{m}{\xi}+1\right)
\Gamma\left(\frac{m}{\xi}\right)}\tlP_m^{-1},
\ee
where $\tlP_m\equiv\tlB_m/\tlA_m$ is the only non-universal
parameter, depending on the 
detailed shape of $V_{\rm 2D}(\rho)$ in the short-distance regime.
Note that above results apply only in the low energy and/or weak interaction
limit, i.e. $k\ll \rho^{-1}\ll \xi^{1/\xi}/\Delta_\alpha$.

For $s$-wave scattering channel, we can define 
an effective scattering length, $a_\alpha\equiv
\Delta_\alpha (2\beta_0\xi)^{\frac{-1}{\xi}}
e^{\frac{-1}{\xi\tlP_0}}$, so that 
\be
\hat{\cal V}_{0}\psi(\bfr_\perp)=
\frac{-2\pi\hbar^2}{2\mu\ln(ka_\alpha/2\beta_0)} 
\delta(\bfr_\perp)\psi(\bfr_\perp),
\label{V_ps_alpha}
\ee
if we assume the scattered wavefunction can be approximated
by a smooth function at origin after cross-graining the short-ranged
fluctuation (for example, the meanfield condensate wavefunction 
of bosonic particles). The resulting pseudo-potential above
becomes the same as a hard-disk potential 
[\oncite{schick,2D_pseudo,review}] with an effective "radius",
$a_\alpha$. The justification of 
the 2D pseudo-potential depends on the interaction strength, i.e.
when $ka_\alpha\sim k\Delta_\alpha\ll 1$.

\begin{figure}
\includegraphics[width=8cm]{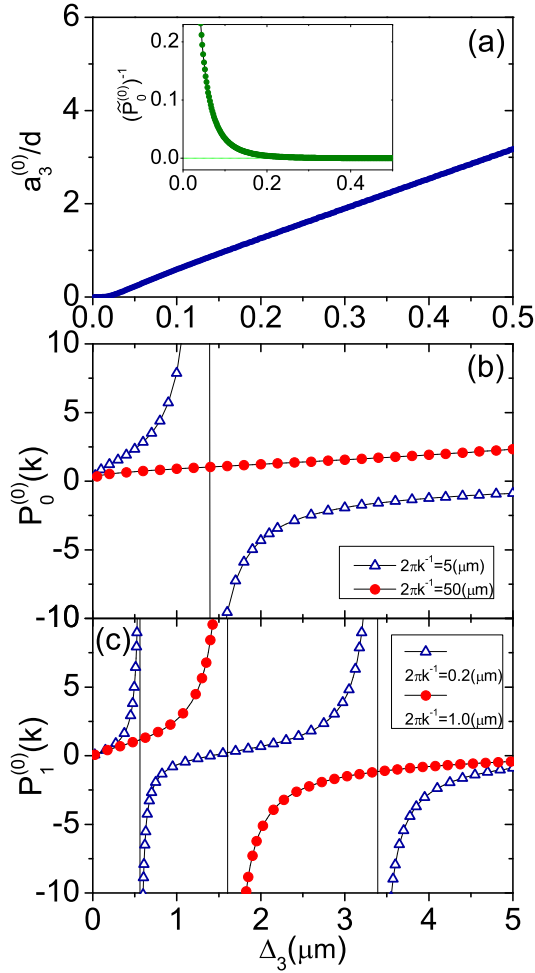}
\caption{
Results for Case A: (a) and the inset are respectively the
value of $a_3^{(0)}$ and $(\tlP_0^{(0)})^{-1}$ as a function 
of $\Delta_3$. Here we set $W=0.1$ $\mu$m (see
the text).  (b) and (c) show the calculated  
${\cal P}^{(0)}_0(k)$ and ${\cal P}^{(0)}_1(k)$ respectively
for different values of incident wavevector, $k$.
}
\label{intralayer}
\end{figure}
\underline{\it Numerical results for the dipolar gases:}
In the rest of this paper, we will concentrate on a physical
example, say systems of polar molecules, for the results of 
dipolar interaction ($\alpha=3$). An external electric field is assumed
to applied perpendicular to the layer 
plane, inducing a field-dependent dipole moment, $D$.
We consider three cases of scattering here: Case A: 
$s$-wave scattering between identical bosons in the same 
layer, Case B: $p$-wave scattering between identical fermions 
in the same layer, and Case C: $s$-wave scattering 
between identical bosons or fermions in two parallel 
layers with layer separation $d$. The last case can be directly
applied to the systems of multi-layer structure made by 1D optical lattice 
[\oncite{chain_wang}]. In the rest of the paper, we
will use $V_{\rm 2D}^{(0)/(1)}$ to denote the bare intra-/inter-layer 
interaction, with the superscript, $^{(0)/(1)}$, to
identify all the quantities obtained by either of them.
For the convenience of numerical
calculation, we further approximate the effective 2D interaction
by the following analytic form: for the intra-layer
interaction, we use $V^{(0)}_{\rm 2D}(\rho)
=\frac{D^2}{\rho^3}$ for $\rho>W$ and $=\frac{D^2}{W^3}$ 
for $\rho\leq W$, where $W$ should be about the same
order of the layer width and is fixed to be 0.1 $\mu$m in the following
calculation. We note that different choices of the cut-off, $W$, 
can bring only minor quantitative difference in the results of phase 
shift (not shown here), because the intra-layer interaction,
$V^{(0)}_{\rm 2D}(\rho)$, is assumed to be repulsive for all $\rho$, and hence
no Feshbach resonance type resonance should be expected.
For the inter-layer interaction, we use
$V^{(1)}_{\rm 2D}(\rho)=\frac{D^2(\rho^2-2d^2)}
{(\rho^2+d^2)^{5/2}}$, where the effect of finite layer width is expected to be
smaller since $W\ll d$ in a deep optical lattice. 
As a result, the only length scale associate with the our repsent model
interaction is $\Delta_3=MD^2/\hbar^2$ (also
denoted to be $a_d$ in the literature [\oncite{Cr}]). For a typical
molecule, say SrO, the fully polarized dipole moment can be
$D=8.9$ Debye, leading to $\Delta_3$ as large as
123.2 $\mu$m. However, for magnetic atoms like $^{52}$Cr,
the maximum value of $\Delta_3$ is just about 1.03 nm.

\begin{figure}
\includegraphics[width=8cm]{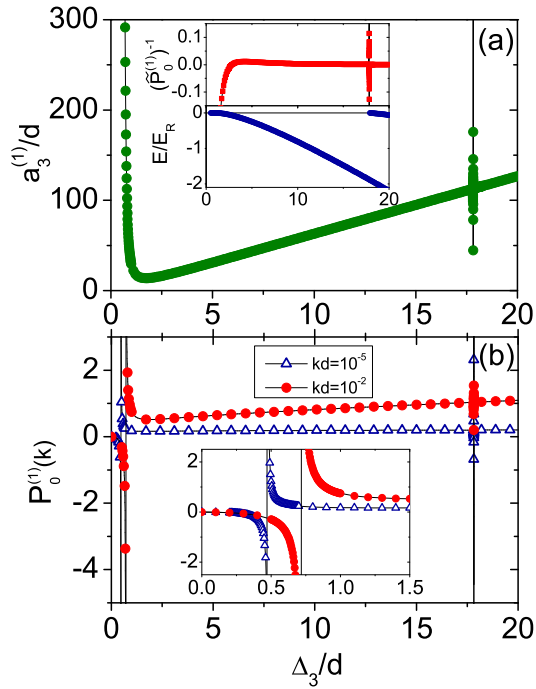}
\caption{
Results for Case C: (a) is the value of $a^{(1)}_3/d$ as a function of
$\Delta_3/d$. In the inset: the upper one shows 
$(\tlP_0^{(1)}{})^{-1}$ for zero energy scattering, and the lower 
one shows the bound state energy in unit of recoil
energy, $E_R$. Note that all the 
length scales are in unit of the inter-layer separation, $d$. (b) shows 
the value of ${\cal P}_0^{(1)}(k)$ for different incident 
wavevectors, $k$. Inset: the magnified plot for the first 
resonance.
}
\label{interlayer}
\end{figure}
In Fig. \ref{intralayer}, we show the numerical results for Case A and 
Case B together by evaluating the original two-particle Schr\"{o}dinger
equation of Eq. (\ref{eq_2body}):
In (a), we show $\tlP_0^{(0)}$ and $a_3^{(0)}$ as a function 
of dipolar strength, $\Delta_3$.
One can see that when $\Delta_3$ is small,
$(\tlP_0^{(0)})^{-1}$ can be quiet large, leading to a very small
scattering length, $a_3^{(0)}$ for $\Delta_3<0.05$ $\mu$m. However,
for larger $\Delta_3$, $a_3^{(0)}$ becomes proportional to $\Delta_3$, 
which is the only
relevant length scale in this regime (i.e. the short-ranged details of
the dipolar interaction becomes negligible).  
In Fig. \ref{intralayer}(b), we show the calculated strength
of pseudo-potential, ${\cal P}^{(0)}_0(k)$, for different values 
of incident wavevectors, $k$. One can see that for a given
$k$, ${\cal P}_0^{(0)}(k)$ decreases to zero logarithmically as $\Delta_3\to 0$, 
while it has a resonance-like behavior at a certain value of $\Delta_3$. 
Such "resonance-like" behavior occurs as $\delta_0(k)=\pi/2$, 
indicating that the interaction is so strong to 
push the wavefront of the scattered wavefunction well-ahead of the 
noninteracting one. It is therefore nothing to do with the Feshbach 
resonance in 3D, and only results for $ka_3\sim k\Delta_3\ll 1$ are
correct for true low energy scattering.
In (b) we show the results for Case B: ${\cal P}_1^{(0)}(k)$ as
a function of dipolar strength. However, unlike the boson case, where 
the typical incident wavevector is determined by the condensate 
(i.e. system) size at low temperature, the typical incident
wavevector for fermions at low temperature should be about
the Fermi wavevector (i.e. the inverse of inter-particle distance) 
due to the Pauli exclusion principle. Therefore we calculate 
results for a much larger $k$ in (c), but similar interaction 
dependence is still observed.

In Fig. \ref{interlayer}, we show the results for Case C:
the $s$-wave scattering between particles
in two different layers. In (a) and its inset, we show 
the calculated scattering length, $a_3^{(1)}$, and
the associated $\tlP_0^{(1)}$ as a function of $\Delta_3$. 
It is interesting to see that, different from the intra-layer case, 
$(\tlP_0^{(1)})^{-1}$ diverges to negative infinity and $a_3^{(1)}$ 
also diverges in the regime of small $\Delta_3$.
Such divergence originates 
from the fact that our dipolar interaction can always sustain 
an inter-layer bound state in 2D system, even when the interaction strength is 
infinitely small. The calculated bound state energy (also in the inset) shows 
a logarithmically small binding energy for the first bound state, 
while the second bound state appears near
$\Delta_3/d\sim 71$. When considering the finite size effect, i.e. $ka$ is bounded
below, the first resonance will occur at a finite dipole moment as shown
in (b) (also see Ref. [\oncite{bilayer_wang}]). The existence of an inter-layer 
bound state can lead to a strong modification of the pseudo-potential 
strength (similar to the Feshbach resonance), leading to some
exotic many-body phases as predicted in Refs. [\oncite{bilayer_wang,chain_wang}].

In summary, we analytically derive the general form of the
pseudo-potential for an arbitrary short-ranged and isotropic 
interaction in a uniform 2D system. 
The energy and interaction dependence of the pseudo-potential 
is also derived analytically for an arbitrary power-law interaction.
Numerical results are provided for 
the dipolar interaction, and therefore can be applied in the  
study of the 2D quantum dipolar gases.

We thank K. Kanjilal and D. Blume for discussion.
This work is supported by NSC and NCTS in Taiwan.


\end{document}